\documentclass[preprint,authoryear,12pt]{elsarticle}

\usepackage{amsfonts}
\usepackage{graphicx, amsmath, amssymb,bm}
\usepackage[english]{babel}

\newcommand{\koff}{k_{\text{off}}}
\newcommand{\kon}{k_{\text{on}}}
\newcommand{\kfb}{k_{\text{fb}}}

\begin{document}

\Large
\centerline{Stochastic pattern formation and spontaneous polarisation:}

\centerline{the linear noise approximation and beyond}
\normalsize

\vspace{1cm}

\centerline{Alan J.~McKane$^1$, Tommaso Biancalani$^1$ and Tim Rogers$^{1,2}$}

\medskip

$^1$Theoretical Physics Division, School of Physics and Astronomy, 
University of Manchester, Manchester M13 9PL, United Kingdom

\medskip

$^2$Department of Mathematical Sciences, University of Bath, Claverton 
Down, Bath BA2 7AY, United Kingdom

\vspace{1cm}

\centerline{\bf Abstract}
\vspace{5mm}

We review the mathematical formalism underlying the modelling of stochasticity
in biological systems. Beginning with a description of the system in terms 
of its basic constituents, we derive the mesoscopic equations governing the
dynamics which generalise the more familiar macroscopic equations. We apply 
this formalism to the analysis of two specific noise-induced phenomena observed
in biologically-inspired models. In the first example, we show how the 
stochastic amplification of a Turing instability gives rise to spatial and 
temporal patterns which may be understood within the linear noise approximation.
The second example concerns the spontaneous emergence of cell polarity, where 
we make analytic progress by exploiting a separation of time-scales. 

\vspace{1.5cm}

\section{Introduction}
\label{intro}
The evidence from the talks delivered at this meeting should leave little 
doubt that stochastic models are required to understand a wide range of 
biological phenomena. The models do need to be carefully formulated, with the
nature of the interactions between the constituents clearly specified, but 
when this is carried out in an unambiguous fashion, their analysis can begin.
In the great majority of cases this will be numerical, and many of the 
contributions at this meeting were concerned with the development and 
implementation of efficient algorithms for carrying this analysis out.

In this paper we will take a different path. We will similarly take care in
formulating the precise form of the model, stressing the need to begin from a
microscopic individual-based model (IBM). However we will be primarily be 
interested in using mathematical analysis to understand the model. To 
facilitate this, the interactions between the constituents of the model will 
be represented by chemical-like reactions occurring at specified rates 
\citep{black12}. If the underlying stochastic process is Markov, which will 
usually be the case, a master equation can be written down which will 
describe the time evolution of the probability of the system being in a 
given state \citep{Kampen97}. 

One analytic procedure that can always be carried out on the master equation 
is to calculate the equations describing the time evolution of the averages
of the state variables. These deterministic equations give a macroscopic
description of the dynamics of the system. They are the other major 
methodology for the theoretical study of biological systems, and their use is
exemplified by the book by \cite{Murray08}. This older tradition 
involved both the study of simple, analytically tractable, models and 
dynamical systems theory. The former was concerned with the mathematical 
investigation of specific differential equations of few variables and the 
latter with general results on stability of attractors, topological notions, 
bifurcation theory, and so on \citep{Wiggins03}.

A parallel methodology for the mathematical analysis of the full stochastic 
model also exists, however it is much less widely appreciated than that for 
the corresponding deterministic analysis. Stochastic differential equations 
(SDEs) can be derived when the number of constituents is large (but not
infinite); the stochasticity originating from the discreteness of the 
underlying IBM. Techniques from the theory of stochastic processes, as well
as general results from this theory, can be used to understand these equations 
analytically, just as in the deterministic case. 

Here we will describe this methodology with reference to two specific models
in order to illustrate the basic ideas and techniques. We will begin in 
section 2 with the definition of the state variables of the model, and the 
specification of the interactions between them in terms of chemical reactions.
This allows us to write down the master equation for the process. In section 
3 we derive the macroscopic and mesoscopic equations governing the dynamics 
of the process from this master equation. We then go on to illustrate the use 
of this formalism, within the linear noise approximation (LNA), to the 
particular example of the Brusselator model in section 4. However, the LNA is 
not always able to capture the details of stochastic ordering phenomena, and 
in section 5 we show that, in some situations, we are able to go beyond the 
LNA. This is illustrated on the spontaneous emergence of cell polarity; this 
example being motivated by the talk given by Linda Petzold at the meeting. 
Finally we conclude with an overview of the techniques and applications that 
we have discussed in section 6.

\section{Definition of the models and the master equation}     
\label{micro}
The stochastic effects that will interest us here occur when the number of
constituents, $N$, is large but finite. In this case a \textit{mesoscopic} 
description is required: the state variables are assumed continuous --- unlike
in the microscopic IBM description --- but the effect of the discreteness is  
not lost: it is manifested in the form of Gaussian distributed `noise'. 
The mesoscopic form of the model is not obvious; we have to begin from a 
microscopic IBM and derive it as an approximation which holds when $N$ is large.
It cannot be found from a macroscopic description, since this consists of 
time evolution equations for average quantities, whereas the micro- or 
mesoscopic formulations describe the time evolution of the entire probability 
density function (pdf). But many stochastic processes have the same average, so
there is in principle no unique way of determining the micro- or mesoscopic 
model from the macroscopic one. This will be especially clear later, when
we derive the mesoscopic form of the equations and compare them to their
macroscopic counterparts.

To set up the IBM we first need to decide what are the variables which describe 
the state of the system. For the simplest case of a single type of 
constituent with no spatial, class, or other structure, it would be a single
integer $n=1,2,\ldots,N$ representing the number of constituents in the system.
In ecological models this could be the number of individuals in the population. 
There would then be a transition rate from state $n$ to state $n'$ caused by 
births, deaths, competition, predation, etc. This rate will be denoted by 
$T(n'|n)$, with the initial state on the left (the other convention with the 
initial state on the right is also sometimes used). 

The probability of finding the system in state $n$ at time $t$, $P_n(t)$, 
changes according to the master equation:
\begin{equation}
\frac{\mathrm{d}P_n(t)}{\mathrm{d}t} = \sum_{n' \neq n}\,T(n|n ')\,P_{n'}(t) 
- \sum_{n' \neq n}\,T(n '|n)\,P_n(t).
\label{master1}
\end{equation}
The first term on the right-hand side describes the rate at which $P_n(t)$ 
increases due to transitions into the state $n$ from all other states $n'$ while
the second term on the right-hand side describes the rate at which $P_n(t)$ 
decreases due to transitions out of the state $n$ to all other states $n'$.
The net change then gives $dP_n(t)/dt$. Although it is intuitively clear, it
can also be derived from the Chapman-Kolmogorov equation for Markov processes
\citep{Gardiner09}. It gives the `microscopic' description of the system and 
will be the starting point for deriving the meso- and macroscopic descriptions.

All this generalises to several types of constituents with numbers 
$\ell, m,\ldots$ at a given time. Spatial structure can also be introduced with
constituents located in a particular small volume $j=1,2\ldots$ at a given 
time. For notational simplicity we will combine the index $i$ labelling these 
small volumes with an index $s$ which labels the types or classes of 
constituent, into one label $J=\{ j,s \}$. Later, when carrying out
the analysis of the differential equations we will separate the indices, and 
may also take the continuum limit in which volume labels become continuous. An 
additional comment is worth making at this stage. There is no agreed 
nomenclature for the small volumes: describing them as `cells' is potentially
confusing in a biochemical context and the term `patches' is usually only
used in an ecological context when the constituents are individuals. We will
use the neutral term `domain' and talk about their `volume' $V$, even when 
they are one- or two-dimensional, as will be the case with the models we
discuss in this paper.  
 
If we now write $n_I$ for the number of constituents of a particular type in a 
particular domain, we can specify the state of the system through the vector 
of integers $\boldsymbol{n} = (n_1, n_2, \ldots)$. The master equation is 
then simply Eq.~(\ref{master1}) with $n$ replaced by $\boldsymbol{n}$:
\begin{equation}
\frac{\mathrm{d}P_{\boldsymbol{n}}(t)}{\mathrm{d}t} = 
\sum_{\boldsymbol{n}' \neq \boldsymbol{n}}\,T(\boldsymbol{n}|\boldsymbol{n}')
P_{\boldsymbol{n}'}(t) -
\sum_{\boldsymbol{n}' \neq \boldsymbol{n}}\,T(\boldsymbol{n}'|\boldsymbol{n})
P_{\boldsymbol{n}}(t). 
\label{master_gen}
\end{equation}

Having decided what the fundamental constituents of the system are, and so 
how the states of the system are defined, the next step is to give the
transition rates $T(\boldsymbol{n}|\boldsymbol{n}')$. This will define the 
model, and is best done through examples. We will naturally choose as examples 
the models that will be analysed later in the paper. The first is the 
Brusselator \citep{Prigogine71} and the second is a model of cell polarity 
introduced by \cite{Altschuler08}.

The notation used to describe the chemical types and the rates in the 
Brusselator model follows that of \cite{Cross09}. In every domain $i$ 
molecules of two species $X$ and $Y$ interact through the reactions of the 
Brusselator model \citep{Prigogine71}:
\begin{eqnarray}
\emptyset &\stackrel{a}{\rightarrow}& X_i, \nonumber \\
X_i &\stackrel{b}{\rightarrow}& Y_i, \nonumber \\ 
2 X_i + Y_i  &\stackrel{c}{\rightarrow}& 3 X_i, \nonumber \\
X_i &\stackrel{d}{\rightarrow}& \emptyset. 
\label{Bruss_reactions_loc}
\end{eqnarray} 
In order, these reactions describe: (i) the creation of a new $X$ molecule, 
(ii) an $X$ molecule spontaneously transforming into a $Y$ molecule, (iii) 
two $X$ molecules reacting with a $Y$, changing it to an $X$, and (iv) $X$ 
molecules being removed from the system. The rates at which the reactions 
occur are denoted by $a,b,c$ and $d$, and $X_i$ and $Y_i$ are the molecules 
that are in domain $i$ at the time that the reaction occurs. Each of these 
reactions are assumed to occur independently, and without memory of the 
previous states of the system. In addition to the reactions given in 
Eq.~(\ref{Bruss_reactions_loc}), migration reactions, which describe molecular
diffusion from one domain to another, have to be specified. For every pair of 
neighbouring domains $i$ and $j$, molecules of the two species $X$ and $Y$ may 
diffuse from one domain to the other according to 
\begin{eqnarray}
X_i \stackrel{\alpha}{\rightarrow} X_j, \ \  \ 
Y_i \stackrel{\beta}{\rightarrow} Y_j\  .
\label{Bruss_reactions_mig}
\end{eqnarray} 
The second example consists of a two-dimensional `cell' enclosed within a 
perfectly circular membrane \citep{Altschuler08}. The centre of the cell is 
known as the `cytoplasmic pool' and contains signalling molecules which we 
denote by $C$. In our implementation, the one-dimensional circular membrane is 
divided up into domains labelled by an index $i$; molecules lying within 
domains $i$ are denoted by $M_i$. The following reactions take place:
\begin{eqnarray}
C &\stackrel{k_{\rm on}}{\rightarrow}& M_i, \nonumber \\
M_i &\stackrel{k_{\rm off}}{\rightarrow}& C, \nonumber \\ 
M_i + C  &\stackrel{k_{fb}}{\rightarrow}& 2 M_i.
\label{polar_reactions_loc}
\end{eqnarray} 
The first two reactions describe a molecule in the cytoplasmic pool attaching 
itself to a random domain on the membrane and a molecule detaching and returning
to the cytoplasmic pool, respectively. The third reaction represents a 
molecule on the membrane attempting to recruit another molecule from the 
cytoplasmic pool to the same domain. As for the Brusselator, there is also 
diffusion, in this case along the membrane:
\begin{eqnarray}
M_i &\stackrel{\alpha}{\rightarrow}& M_j\,,
\label{polar_reactions_mig}
\end{eqnarray} 
for any pair of neighbouring membrane domains $i$ and $j$. The reaction 
scheme is illustrated in Fig.~1 (after Altschuler \textit{et al}).


\begin{figure}[t]
\begin{center}
\includegraphics[width=0.45\textwidth, trim=20 0 100 450]{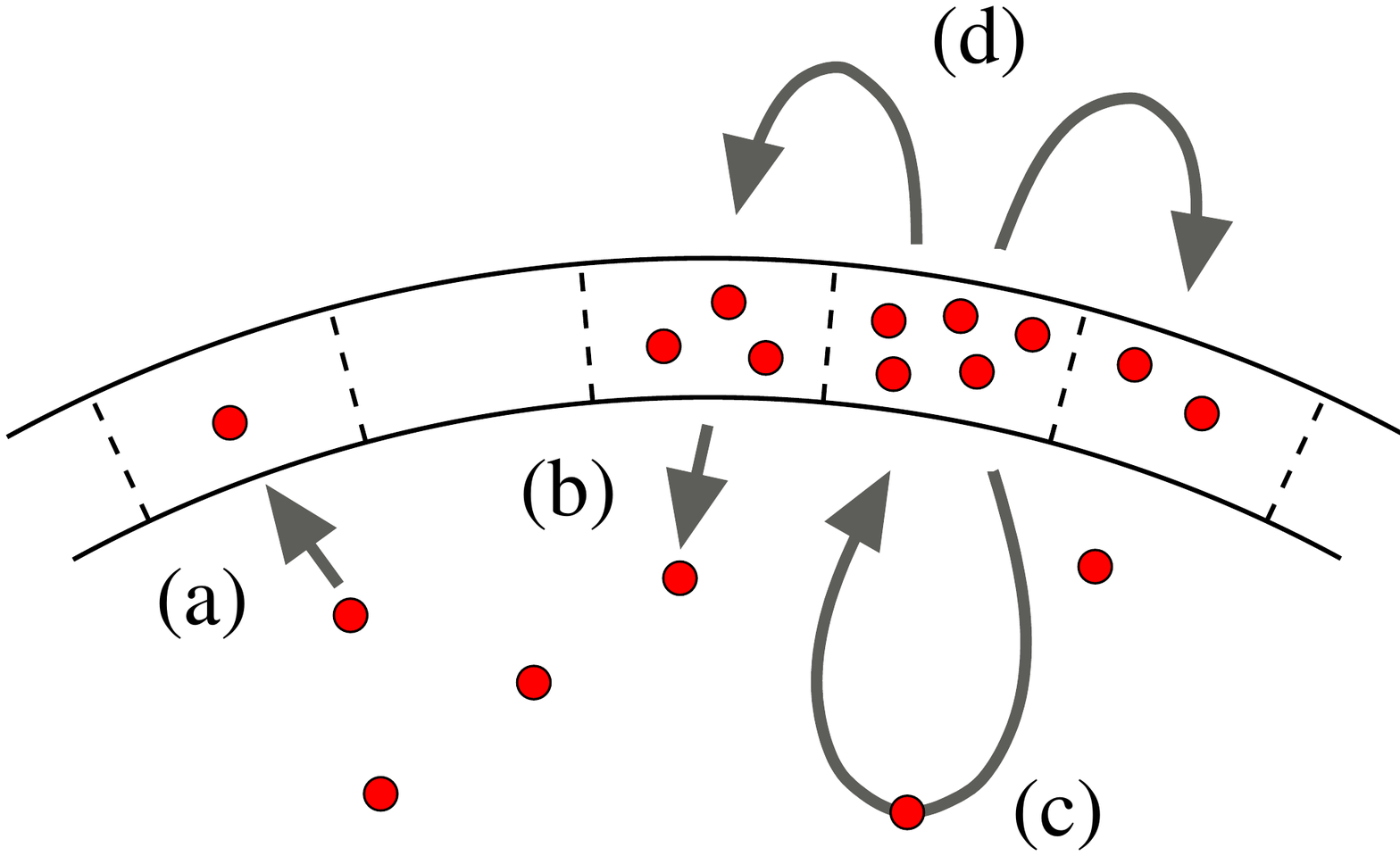}
\caption{Illustration of the reactions of the cell polarisation model. The three
reactions are (a) attachment of signalling molecules to the membrane, (b) the
return of a membrane molecule to the cytoplasmic pool, (c) recruitment of new
signalling molecules by those already on the membrane, and (d) diffusion along
the membrane, which we implement as migration between neighbouring domains.}
\label{fig:cell}
\end{center}
\end{figure}


Once the reaction scheme has been picked and laid out as in 
Eqs.~(\ref{Bruss_reactions_loc})-(\ref{polar_reactions_mig}), the transition
rates $T(\boldsymbol{n}|\boldsymbol{n}')$ can be chosen. This effectively 
specifies the model. When writing the transition rates, we will only list the 
variables in the domains that are involved in the reaction.

For the Brusselator, the number of $X_i$ and $Y_i$ will be denoted by $\ell_i$ 
and $m_i$, respectively. We then invoke mass action, that is, the rate of the 
transition for a given reaction is proportional to the product of the 
densities of the reactants, to write the transition rates as
\begin{eqnarray}
T_{1}(\ell_i + 1,m_i|\ell_i,m_i) &=& a, \nonumber \\
T_{2}(\ell_i-1,m_i+1|\ell_i,m_i) &=& b \; \frac{\ell_i}{V}, \nonumber \\
T_{3}(\ell_i+1, m_i-1|\ell_i,m_i) &=& c \; \frac{{\ell_i}^2m_i}{V^3} \nonumber
\\
T_{4}(\ell_i-1, m_i|\ell_i,m_i) &=& d \; \frac{\ell_i}{V},
\label{Bruss_transprob}
\end{eqnarray}
where the subscripts on the rates refer to the four reactions in 
Eq.~(\ref{Bruss_reactions_loc}) and where $V$ is the volume of the domain. 
These transition rates are as in the usual Brusselator model
\citep{Boland09,Biancalani11}, although it is worth noting that we have not 
imposed a fixed limit on the number of molecules permitted in a domain (in 
contrast with some previous work~\citep{Lugo08,Biancalani10}; this is 
reflected in the fact we use the inverse volume of a domain, $V^{-1}$, rather 
than the inverse total number of molecules in a domain, as the expansion 
parameter). 

The reaction rates which describe molecular diffusion reactions
in Eq.~(\ref{Bruss_reactions_mig}) are given by
\begin{eqnarray}
T_{5}(\ell_i-1, \ell_j+1|\ell_i,\ell_j) &=& \alpha \; \frac{\ell_i}{zV}, 
\nonumber \\
T_{6}(m_i-1,m_j+1|m_i,m_j) &=& \beta \; \frac{m_i}{zV}.
\label{Bruss_migprob}
\end{eqnarray}
Here $z$ is the number of nearest neighbours of a domain and the index $j$ 
denotes a nearest neighbour of domain $i$. 

To model spontaneous cell polarisation we denote the number of $C$ and $M_i$ 
molecules by $\ell$ and $m_i$, respectively. The transition rates are then 
taken to be
\begin{eqnarray}
T_{1}(\ell - 1,m_i+1|\ell,m_i) &=& k_{\rm on} \; \frac{\ell}{V} \nonumber \\
T_{2}(\ell + 1,m_i-1|\ell,m_i) &=& k_{\rm off} \; \frac{m_i}{V}, \nonumber \\
T_{3}(\ell -1, m_i+1|\ell,m_i) &=& k_{\rm fb} \; \frac{\ell m_i}{V^2},
\label{polar_transprob}
\end{eqnarray}
for the reactions (\ref{polar_reactions_loc}) and 
\begin{eqnarray}
T_{4}(m_i-1, m_j+1|m_i,m_j) &=& \alpha \; \frac{m_i}{2V}, 
\label{polar_migprob}
\end{eqnarray}
for the reactions (\ref{polar_reactions_mig}).

The transition rates for these two models can be substituted into 
Eq.~(\ref{master_gen}) which can then, together with suitable initial and
boundary conditions, be used to solve for $P_{\boldsymbol{n}}(t)$. They can
also 
be used as the basis for setting up a simulation using the Gillespie 
algorithm \citep{Gillespie76,Gillespie77}. Our approach in this paper will be 
to devise approximation schemes for the master equation and to check the 
results obtained from such schemes with simulations based on the Gillespie 
algorithm.


We end this section by generalising the above formulation so that it applies 
to a general biochemical network and by extension to any network of interacting
agents. To do this, suppose that there are $L$ different constituents in the 
system. These could be labelled by their type, the domain they occupy, etc. We 
will denote them by $Z_I, I=1,\ldots,L$ and at a given time there will be 
$n_I$ of them, so that the state of the system can be specified by  
$\boldsymbol{n} = (n_1,\ldots,n_L)$, as described earlier in this section. We
suppose that there are $M$ reactions which interconvert species:
\begin{equation}
\sum_{I=1}^{L} r_{I \mu}Z_I \ {\longrightarrow}
\ \sum_{I=1}^{L} p_{I \mu} Z_I, \ \ \ \mu=1,2,...M.
\label{gen_reaction}
\end{equation}
Here the numbers $r_{I \mu}$ and $p_{I \mu}\ (I=1,\ldots,L;\mu=1,\ldots,M)$ 
describe respectively the population of the reactants and the products 
involved in the reaction. The reactions
Eqs.~(\ref{Bruss_reactions_loc})-(\ref{polar_reactions_mig}) are simple 
examples of this general set of reactions.

A quantity which is central to the structure of both the mesoscopic and 
macroscopic equations is the stoichiometry matrix, $\nu_{I \mu}
\equiv r_{I \mu}-p_{I \mu}$, which describes how many molecules of
species $Z_I$ are transformed due to the reaction $\mu$. In the notation 
introduced above for the master equation, 
$\boldsymbol{n}' = \boldsymbol{n}-\boldsymbol{\nu}$, where 
$\boldsymbol{\nu}_{\mu} = (\nu_{1 \mu},\ldots,\nu_{L \mu})$ is the
stoichiometric vector corresponding to reaction $\mu$. Therefore the master 
equation (\ref{master_gen}) may be equivalently written as
\begin{equation}
\frac{\mathrm{d}P_{\boldsymbol{n}}(t)}{\mathrm{d}t} = \sum^{M}_{\mu=1}
\left[ T_{\mu}(\boldsymbol{n}|\boldsymbol{n}-\boldsymbol{\nu}_{\mu})
P_{\boldsymbol{n}-\boldsymbol{\nu}_{\mu}}(t) - 
T_{\mu}(\boldsymbol{n}+\boldsymbol{\nu}_{\mu}|\boldsymbol{n})
P_{\boldsymbol{n}}(t) \right].
\label{master_alt}
\end{equation}
Many models of interest involve only a handful of different reactions; in 
this situation, it is often convenient to rewrite the master equation 
as a sum over reactions as in Eq.~(\ref{master_alt}), rather than over pairs 
of states $\bm{n}$ and $\bm{n}'$ as in Eq.~(\ref{master_gen}). In the next 
section we will describe how the mesoscopic description of the models can be 
obtained from the master equation (\ref{master_alt}). These can then be used 
as the basis for the calculational schemes which we wish to implement.

\section{Derivation of the macro- and mesoscopic equations}     
\label{meso}
\subsection{Macroscopic equation}
Before we derive the mesoscopic equations from the master equation, we will 
carry out the simpler task of deriving the macroscopic equations. This will be
done for the case of the general biochemical network described in section 
\ref{micro}.

This is achieved by multiplying Eq.~(\ref{master_alt}) by $\boldsymbol{n}$, 
and summing over all possible values of $\boldsymbol{n}$. After making the
change of variable $\boldsymbol{n} \rightarrow
\boldsymbol{n}+\boldsymbol{\nu}$ in the first summation, one finds that
\begin{equation}
\frac{\mathrm{d}\langle \boldsymbol{n}(t) \rangle}{\mathrm{d}t} = 
\sum^{M}_{\mu=1} \boldsymbol{\nu}_{\mu} 
\big\langle T_{\mu}(\boldsymbol{n}+\boldsymbol{\nu}_{\mu}|\boldsymbol{n}) 
\big\rangle,
\label{av_n_eqn}
\end{equation}
where the angle brackets define the expectation value:
\begin{equation}
\langle\cdots\rangle = \sum_{\bm{n}}(\cdots)P_{\bm{n}}(t)\,.
\end{equation} 

In the limit where both the particle numbers and the volume become large, we
will take the state variables to be the concentration of the constituents
$n_I/V$, rather than their number $n_I$. These will be assumed to have a 
finite limit as $V \to \infty$. Specifically, the state of the system will be
determined by the new variables 
$$y_I=\lim_{V \to \infty} \frac{\langle n_I\rangle}{V}\,,\quad\text{where}\quad
I=1,\ldots,L\,.$$ 
From Eq.~(\ref{av_n_eqn}) we have that
\begin{equation}
\frac{\mathrm{d}y_I}{\mathrm{d}\tau}=
\sum_{\mu=1}^M \nu_{I \mu}f_{\mu}(\boldsymbol{y}), 
\ \ \ I=1,\ldots,L,
\label{ODEs_gen}
\end{equation}
where $\tau = t/V$ and where
\begin{eqnarray}
f_{\mu}(\boldsymbol{y})&=&\lim_{V \to \infty} \big\langle
T_{\mu}(\boldsymbol{n}+\boldsymbol{\nu}_{\mu}|\boldsymbol{n}) 
\big\rangle\bigg.\nonumber\\
&=& \lim_{V \to \infty} T_{\mu}\big(\langle\boldsymbol{n}\rangle+
\boldsymbol{\nu}_{\mu}|\langle\boldsymbol{n}\rangle\big) \bigg.\nonumber\\
&=&\lim_{V \to \infty} T_{\mu}(V\boldsymbol{y}+
\boldsymbol{\nu}_{\mu}|V\boldsymbol{y})\bigg.\,.
\label{f_defn}
\end{eqnarray}
In the above we have used the fact that in the macroscopic limit the 
probability distribution functions are Dirac delta functions and so, for 
instance, $\langle n^m \rangle = \langle n \rangle^m$, for any integer $m$. 

The equation 
\begin{equation}
\frac{\mathrm{d}y_I}{\mathrm{d}\tau} = A_I(\boldsymbol{y}),
\label{macroscopic_eqn}
\end{equation}
where
\begin{equation}
A_I(\boldsymbol{y}) \equiv \sum_{\mu=1}^M \nu_{I \mu}f_{\mu}(\boldsymbol{y}), 
\ \ \ I=1,\ldots,L,
\label{A_I}
\end{equation}
is the macroscopic equation corresponding to the microscopic master equation
(\ref{master_alt}). It can be calculated from a knowledge of the stoichiometric
matrix $\nu_{I \mu}$ and the transition rates 
$T_{\mu}(\boldsymbol{n}+\boldsymbol{\nu}_{\mu}|\boldsymbol{n})$. We scaled time 
by a factor of $V$ simply because the choice we made for the transition rates
(\ref{Bruss_transprob})-(\ref{polar_migprob}) were finite as $V \to \infty$,
but we could have easily incorporated an extra factor of $V$ in these rates 
through a time rescaling. We also chose particularly simple forms for these 
transition rates in that they were all functions of the species concentration
$n_I/V$. More generally, they might separately be functions of $n_I$ and $V$, 
which become functions of the species concentration $n_I/V$ only when both the 
particle numbers and the volume become large, so that in the limit 
$V \to \infty$ they become functions of the macroscopic state variable 
$\boldsymbol{y}$.

\subsection{Mesoscopic equation}
It is perhaps useful at this stage to recall precisely what is meant by the
terms `microscopic', `mesoscopic' and `macroscopic'. The microscopic 
description is the one based on the fundamental constituents whose reactions 
are described by relations such as those in
Eqs.~(\ref{Bruss_reactions_loc})-(\ref{polar_reactions_mig}). The dynamics of
the processes are described by the 
master equation or by the Gillespie algorithm. The macroscopic description 
has been derived from this microscopic description above: it only involves 
average quantities and their time evolution. In between these two levels of
description is the `meso'-level description, where the 
continuous variable of the macro-description is used, but where the stochastic
effects due to the discrete nature of the individuals is retained. Some other
authors include master equations at the meso-level leaving the micro-level 
for the world of atoms and molecules, but this does not seem such a useful
assignment in the biological context in which we are working. The derivation 
of the mesoscopic equation follows similar lines to the calculation above, 
with the important difference that we do not take an average, or equivalently,
do not take the limit $V \to \infty$.

We begin by substituting $y_I = n_I/V$ directly into the master equation. 
Since, as discussed above, our transition rates are all functions of $n_I/V$ 
we simply replace 
$T_{\mu}(\boldsymbol{n}+\boldsymbol{\nu}_{\mu}|\boldsymbol{n})$ by
$f_{\mu}(\bm{y})$ in the notation of Eq.~(\ref{f_defn}). In addition
we will denote the pdf $P_{\boldsymbol{n}}(t)$ where $\boldsymbol{n}$ has been 
replaced by $V \bm{y}$ as $P(\bm{y},t)$. With these changes we may write the 
master equation (\ref{master_alt}) as
\begin{equation}
\frac{\partial P(\bm{y},t)}{\partial t} = 
\sum^{M}_{\mu=1}\bigg[ f_{\mu}\Big(\bm{y} - \frac{\boldsymbol{\nu}_\mu}{V}\Big) 
P\Big(\bm{y} - \frac{\boldsymbol{\nu}_\mu}{V},t\Big)  -  
f_{\mu}(\bm{y}) P(\bm{y},t) \bigg] \nonumber
\end{equation}
For $V$ large, the steps $\bm{\nu}_\mu/V$ are likely to be very small, 
suggesting that we may expand the functions $P$ and $f$ as Taylor series 
around $\bm{y}$. Truncating at order $\mathcal{O}(V^{-2})$, we arrive at 
\begin{eqnarray}
\label{master_expansion}
\frac{\partial P(\bm{y},\tau)}{\partial \tau} &=& 
- \sum^{M}_{\mu=1} \sum_{I} \nu_{I \mu} \frac{\partial }{\partial y_I} 
\left[ f_{\mu}(\bm{y}) P(\bm{y},\tau) \right] \\
&+& \sum^{M}_{\mu=1} \frac{1}{2V} \sum_{I,J} \nu_{I \mu} \nu_{J \mu}
\frac{\partial^2 }{\partial y_I \partial y_J} \left[ f_{\mu}(\bm{y}) 
P(\bm{y},\tau) \right]\,,\nonumber 
\end{eqnarray}
where as before we have absorbed a factor of $V$ into the rescaled time 
variable $\tau=t/V$. This is a Fokker-Planck equation which can be cast into 
the standard form \citep{Risken89, Gardiner09}
\begin{equation}
\frac{\partial P(\bm{y},\tau)}{\partial \tau} = 
- \sum_{I} \frac{\partial }{\partial y_I} 
\left[ A_{I}(\bm{y}) P(\bm{y},\tau) \right] 
+ \frac{1}{2V} \sum_{I,J} \frac{\partial^2 }{\partial y_I \partial y_J} 
\left[ B_{I J}(\bm{y}) P(\bm{y},\tau) \right],
\label{FPE}
\end{equation}
where $A_I(\mathbf{y})$ is defined by Eq.~(\ref{A_I}) and where
\begin{equation}
B_{I J}(\bm{y}) =\sum^{M}_{\mu=1} \nu_{I \mu} \nu_{J \mu}
f_{\mu}(\bm{y}), \ \ I,J=1,\ldots,L.
\label{B_IJ}
\end{equation}

In the Fokker-Planck equation (\ref{FPE}), the continuous nature of the state 
variables indicates that the individual nature of the constituents has been 
lost. However, the stochasticity due to this discreteness has not: it now 
manifests itself through the function $B_{I J}(\bm{y})$. We can see 
this is the case through the presence of the factor $1/V$.

One might ask if this approach is consistent with the previous macroscopic 
derivation. As $V \to \infty$, the Fokker-Planck equation reduces to the 
Liouville equation
\begin{equation}
\frac{\partial P(\bm{y},\tau)}{\partial \tau} = 
- \sum_{I} \frac{\partial }{\partial y_I} 
\left[ A_{I}(\bm{y}) P(\bm{y},\tau) \right].
\label{Liouville}
\end{equation}
One can check by direct substitution that the solution to this equation is
$P(\bm{y},\tau) = \delta(\bm{y}(\tau) - \bm{y} )$ where 
$\delta$ is the Dirac delta function and where $\bm{y}(\tau)$ is the
solution of the macroscopic system (\ref{macroscopic_eqn}); see 
\citep{Gardiner09} for details. 

It is also natural to ask if it is useful to include higher order terms in 
$V^{-1}$. There are sound mathematical reasons for not going to higher order, 
for instance the pdf may become negative \citep{Risken89}. As we will see, 
for the problems that we are interested in here (and many others) very good 
agreement with simulations can be found by working with the Fokker-Planck 
equation (\ref{FPE}). 


The Fokker-Planck equation (\ref{FPE}) provides a mesoscopic description of the
system but, like the master equation (\ref{master_gen}) from which it 
originated, it is an equation for a pdf. It is therefore quite distinct from 
the macroscopic equation (\ref{macroscopic_eqn}), which is an equation for the
state variables themselves. There does, however, exist an equation for the
state variables which is completely equivalent to the Fokker-Planck equation
(\ref{FPE}) \citep{Gardiner09}. This equation takes the form
\begin{equation}
\frac{\mathrm{d}y_I}{\mathrm{d}\tau} = A_I(\bm{y}) +
\frac{1}{\sqrt{V}}\sum_{J} g_{I J}(\bm{y}) \eta_J(\tau),
\label{mesoscopic_eqn}
\end{equation}
where the $\eta_J(\tau)$ are Gaussian white noises with zero mean and correlator
\begin{equation}
\langle \eta_I(\tau) \eta_J(\tau') \rangle = \delta_{I J}
\delta(\tau - \tau'),
\label{correlator}
\end{equation}
and where $g_{I J}(\bm{y})$ is related to $B_{I J}(\bm{y})$ by
\begin{equation}
B_{I J}(\bm{y}) = \sum_{K} g_{I K}(\bm{y})
g_{J K}(\bm{y}).
\label{B_ggT}
\end{equation}

The mesoscopic equation (\ref{mesoscopic_eqn}) generalises the macroscopic 
ordinary differential equation (\ref{macroscopic_eqn}) with the addition of 
noise terms $\boldsymbol{\eta}(\tau)$ and so is a stochastic differential 
equation (SDE). As we will discuss below we need to specify that it is to be 
interpreted in the sense of It\={o} \citep{Gardiner09}. Notice the direct 
relationship between this SDE and the macroscopic ODE: sending 
$V \to \infty$ in Eq.~(\ref{mesoscopic_eqn}) immediately yields equation 
(\ref{macroscopic_eqn}).

It is important to point out that the matrices $g_{IJ}(\bm{y})$ which define 
the behaviour of the noise cannot be found from the macroscopic equations,
and a knowledge of the microscopic stochastic dynamics is essential if one is 
to understand the effects of noise. It is not permissible in this context to 
simply `add noise terms' to the macroscopic equations to obtain a mesoscopic 
description, as some authors have done in the past. The only situation in 
which it is permissible to do this is if the noise is external to the 
system, that is, it does not originate from the internal dynamics of the 
system.

We end this section with two general comments on the mesoscopic equation
(\ref{mesoscopic_eqn}). The first is that while there are no strong 
restrictions on the form of $A_I(\bm{y})$, there are on 
$B_{I J}(\bm{y})$. From Eq.~(\ref{B_IJ}) we see that the matrix $B$ is
symmetric, but also that for any non-zero vector $\boldsymbol{w}$,
\begin{equation}
\sum_{I,J} w_{I} B_{I J} w_{J} = \sum^{M}_{\mu = 1} \left( \boldsymbol{w}\cdot
\boldsymbol{\nu} \right)^{2}f_{\mu}(\bm{y}) \geq 0,
\label{pos_defn}
\end{equation}
since $f_{\mu}(\bm{y}) \geq 0$. Thus $B$ is positive semi-definite 
\citep{Mehta89}. It follows that $B=g\,g^{\rm T}$ for some non-singular 
matrix $g$, where T denotes transpose \citep{Mehta89}. One way of 
constructing such a matrix is to note that since $B$ is symmetric, it can be 
diagonalised by an orthogonal transformation defined through a matrix 
$O_{I J}$. Then since $B$ is positive semi-definite, its eigenvalues are 
non-negative, and so
\begin{equation}
B = O \Lambda O^{\rm T} = g\,g^{\rm T}, \ \ {\rm where\ } g = O \Lambda^{1/2},
\label{decomposition}
\end{equation}
and where $\Lambda$ and $\Lambda^{1/2}$ are the diagonal matrices with 
respectively the eigenvalues and square root of the eigenvalues of $B$ as 
entries. We can take the positive roots of the eigenvalues without loss if
generality, since the sign can always be absorbed in the $\eta_{J}$ factor in
Eq.~(\ref{mesoscopic_eqn}) (its distribution is Gaussian and so invariant 
under sign changes). It should also be pointed out that we can go
further and make an orthogonal transformation on the noise, 
$\zeta_J = \sum_I S_{I J}\eta_J$, and leave Eq.~(\ref{correlator}), and so its 
distribution, unchanged. The transformation matrix $S$ can then be used to 
define a new matrix $G_{I J} = \sum_K g_{I K} S_{J K}$, so that the form of the 
mesoscopic equation (\ref{mesoscopic_eqn}) is unchanged. So while the 
procedure outlined above gives us a way of constructing 
$g_{I J}(\bm{y})$ from $B_{I J}(\bm{y})$, it is not unique. 

The second comment relates to the statement made earlier, that 
Eq.~(\ref{mesoscopic_eqn}) is to be interpreted in the It\={o} sense. The 
singular nature of white noise means that in some cases SDEs are not uniquely 
defined by simply writing down the equation, but have to be supplemented with 
the additional information on how the singular nature of the process is to be 
interpreted \citep{Kampen97,Gardiner09}. This happens when $g_{IJ}$ depends 
on the state of the system $\bm{y}$; the noise is then said to be
multiplicative. As we will see in the next section, this subtlety is not 
relevant within the LNA, since there the $g_{I J}$ is evaluated at a fixed 
point of the dynamics, and so ceases to depend on the state of the system. 
However, when going beyond the LNA, it is an important consideration. If one 
wishes to manipulate a multiplicative noise SDE like 
Eq.~(\ref{mesoscopic_eqn}), then one must employ modified rules of calculus 
which take into account the contribution from the noise. We refer 
to \cite{Gardiner09} for details, and a complete discussion on the 
relationship between the It\={o} formulation and the alternatives. 


This completes our general discussion of the derivation and form of the 
mesoscopic equation. In the next two sections we will apply it to the two 
models which we introduced in Section \ref{micro}. In the first case we will
consider use of the LNA is sufficient for our requirements, but in the 
second case we have to go beyond the LNA. 

\section{Stochastic patterns in population models}     
\label{bruss}
One of the reasons why deterministic reaction-diffusion systems are interesting 
is the fact that they may give rise to ordered structures, either in space or 
time. Many models displaying several different kinds of patterns have been 
extensively discussed in the literature \citep{Murray08, Cross09}. However, the 
mathematical mechanisms which are responsible for the pattern formation are 
few and universal, and they can be conveniently analysed using the simplest 
models. Perhaps the most famous of these mechanisms was put forward by Turing 
in his pioneering study of morphogenesis \citep{Turing52}, and it is now 
referred to as the `Turing instability' \citep{Murray08, Cross09}. It is 
invoked as a central paradigm in various areas of science to explain the 
emergence of steady spatial structures which typically look like `stripes', 
`hexagons' or `spots' \citep{Murray08, Cross09}. 

In this section, we will be interested in reaction-diffusion systems exhibiting
a Turing instability which are composed of discrete entities as described in 
section \ref{micro}. The intrinsic noise in the system will render it 
stochastic. As we shall see, by means of the LNA, one is able to make 
analytical progress and so clarify the role of demographic noise in the 
pattern formation process. We shall show that systems of this kind display 
`stochastic patterns' in addition to the conventional `Turing patterns'. It 
has been suggested that stochastic patterns are responsible for the robustness 
of the patterning observed in population systems
\citep{Butler09, Butler11, Biancalani10} and they have been applied in several
ecological models \citep{Butler09, Bonachela12}, in the dynamics of 
hallucination \citep{Butler12} and in a biological model with stochastic growth
\citep{Woolley11a}. In a similar way, the emergence of stochastic travelling
waves has been studied \citep{Biancalani11}, which has found application in a 
marine predator-prey system \citep{Delius10}. There is also an existing 
literature on stochastic patterning in arid ecosystems \citep{RidolfiBook} 
where the origin of the noise is extrinsic rather than intrinsic.

The following analysis employs the Brusselator model to exemplify the general 
theory. This is a reaction scheme introduced by Lefever and Prigogine in the 
1960s \citep{Prigogine71} as a model of biochemical reactions which showed 
oscillatory behaviour. For our purposes, its interest lies in the fact that 
its spatial version is one of the simplest models which exhibits a Turing 
instability. 

Before we begin the analysis of this model we need to discuss some aspects of
the notation we will use. As explained in section \ref{micro} the labels 
that we have been using so far combine a spatial index with an index for the
type of constituent, for example $J=\{ j,s \}$. This was done in order to 
reduce the clutter of indices. However in the analysis below we will need to 
separate them, since we will assume a regular lattice structure for the 
domains, which will allow us to use Fourier analysis to effectively diagonalise
the spatial part of the system. The Fourier components of a given function 
will be labelled through the argument of that function, leaving only the index 
(e.g.~$s$) labelling the type. Specifically we will choose the spatial 
structure to be a regular $D$-dimensional hypercubic lattice with periodic 
boundary conditions and domain length $l$. Following the conventions 
of \citep{ChaikinBook}, the discrete spatial Fourier transform is then defined 
as:
\begin{equation}  
\label{spatialFT}
\tilde{f}_k = l^D\sum_{j=1}^\Omega e^{-i l k \cdot j}f_j, \ {\rm with\ } \ 
f_j = l^{-D}\Omega^{-1}\sum_{k=1}^\Omega e^{i l k \cdot j}\tilde{f}_k,
\end{equation}
where $\Omega$ is the number of lattice points, $j$ is a $D$-dimensional 
spatial vector and $k$ is its Fourier conjugate. Note that $i$ here is
imaginary 
unit and not a spatial index, and although both $j$ and $k$ are $D$-dimensional 
vectors, we do not explicitly show this, in line with the notation adopted in 
section \ref{micro}. 

We begin the analysis by deriving the well-known deterministic Brusselator 
equations from this microscopic description using Eqs.~(\ref{macroscopic_eqn})
and (\ref{A_I}):
\begin{eqnarray} \label{Bruss_odes}
\frac{\mathrm{d}u_i}{\mathrm{d}\tau} &=& a - \left( b+d \right) u_i + 
cu^2_i v_i + \alpha\Delta u_i, \nonumber \\
\frac{\mathrm{d}v_i}{\mathrm{d}\tau} &=& b u_i - cu^2_i v_i + \beta\Delta v_i,
\end{eqnarray}
where $u_i = \ell_i/V$ and $v_i = m_i/V$, where $u$ is the density of chemical 
species $X$ and $v$ is the density of chemical species $Y$. The symbol $\Delta$ 
represents the discrete Laplacian operator $\Delta f_j = (2/z)\,\sum_{j' \in
\partial j} \left(f_j - f_{j'}\right)$ where  $j' \in \partial j$ indicates that
the domain $j'$ is a nearest neighbour of the domain $j$ and $z$ is the 
co-ordination number of the lattice. The spatial Fourier transform of the 
discrete Laplacian operator reads \citep{Lugo08}:
\begin{equation} \label{FT_Laplacian}
\tilde{\Delta}_{k} = \frac{2}{D}\sum^{D}_{s = 1}
\left[ \cos\left( k_s \,l\right) - 1 \right].
\end{equation}

It is possible to obtain a continuous spatial description, in the deterministic
limit, by taking the limit of small domain length scale, $l\rightarrow 0$
\citep{Lugo08}. 
By doing so, one can recover the traditional partial differential equations 
for reaction-diffusion systems:
\begin{eqnarray}
\frac{\partial u}{\partial \tau} &=& a - \left( b+d \right) u + cu^2 v + 
D_1\nabla^{2}u, \nonumber \\
\frac{\partial v}{\partial \tau} &=& b u - cu^2 v + D_2\nabla^{2}v,
\label{Bruss_pdes}
\end{eqnarray}
where $D_1$ and $D_2$ are obtained by scaling the diffusivities $\alpha$ and 
$\beta$ according to:
\begin{equation} \label{diff_constants}
\frac{1}{2D} l^2 \alpha \mapsto D_1, \ \ \frac{1}{2D} l^2 \beta \mapsto D_2.
\end{equation}
However, we shall keep the space discrete in the following analysis because 
the theory is simpler to describe and it is most convenient for carrying 
out stochastic simulations. We shall also set $l=1$, since this simply amounts 
to a choice of length scale, and this is the simplest choice.

The macroscopic equations (\ref{Bruss_odes}) are Eq.~(\ref{macroscopic_eqn}) 
for the particular case of the Brusselator model. To find the corresponding
mesoscopic equations we need to find the particular form of
Eq.~(\ref{mesoscopic_eqn}) for the Brusselator. This we can do by 
calculating $B_{I J}(\boldsymbol{y})$, defined in Eq.~(\ref{B_IJ}), but we will
find that we do not need to utilise the non-linear equation 
(\ref{mesoscopic_eqn}) to take the fluctuations into account; it is sufficient 
to use only a linearised form. This is the LNA, and is implemented by writing
\begin{equation}
y_{I}(t) = \langle y_{I}(t) \rangle + \frac{\xi_{I}(t)}{\sqrt{V}},
\label{LNA}
\end{equation}
where $\langle y_{I}(t) \rangle$ satisfies the macroscopic equation 
(\ref{macroscopic_eqn}). Substituting Eq.~(\ref{LNA}) into 
Eq.~(\ref{mesoscopic_eqn}), we expand in powers of $1/\sqrt{V}$. The terms 
which are proportional to $1/\sqrt{V}$ give an equation for $\xi_I$:
\begin{equation}
\frac{\mathrm{d}\xi_I}{\mathrm{d}\tau} = \sum_{J} 
{\mathcal J}_{IJ}(\langle \boldsymbol{y} \rangle)\xi_{J} +
\sum_{J} g_{I J}(\langle \boldsymbol{y} \rangle) \eta_J(\tau),
\label{linearised_eqn}
\end{equation}
where ${\mathcal J}$ is the Jacobian of the system. 

In many situations, including the one we are describing here, we are only 
interested in the fixed points of the macroscopic equation, in which case 
$\langle \boldsymbol{y} \rangle = \boldsymbol{y}^{*}$ and the matrices 
${\mathcal J}$ and $g$ can be replaced by their values at the fixed point
$\boldsymbol{y}^{*}$. The SDE (\ref{linearised_eqn}) now involves only 
constant matrices:
\begin{equation} \label{linearised_constant}
\frac{\mathrm{d}\xi_I}{\mathrm{d}\tau} = \sum_{J} 
{\mathcal J}^{*}_{IJ}\xi_{J} + \sum_{J} g^{*}_{I J}\eta_J(\tau).
\end{equation}
For the specific case of the Brusselator the index $I$ includes the spatial
index $i$ and an index $s=1,2$ which distinguishes between the variables 
$u$ and $v$. If we take the spatial Fourier transform of 
Eq.~(\ref{linearised_constant}), translational invariance implies that the
matrices ${\mathcal J}^{*}$ and $g^*$ are diagonalised in the spatial 
variables, and so this equation becomes
\begin{equation} \label{Fourier_linearised}
\frac{\partial \tilde{\xi}_{\gamma}(k, \tau)}{\partial \tau} = 
\sum_{\delta = 1}^{2}{\mathcal J}^{*}_{\gamma \delta}(k)\,\tilde{\xi}_{\delta}
(k, \tau) + \sum_{\delta = 1}^{2} g^{*}_{\gamma
\delta}(k)\,\tilde{\eta}_{\delta}
(k, \tau).
\end{equation}

We are now in a position to discuss both the classical Turing patterns found
in deterministic equations such as Eqs.~(\ref{Bruss_pdes}) and the stochastic
Turing patterns found in the corresponding mesoscopic equations. The 
homogeneous fixed point of Eqs.~(\ref{Bruss_pdes}) is given by 
\begin{equation}
u^* = \frac{a}{d}, \ \ v^* = \frac{bd}{ac},
\label{FP}
\end{equation}
although from now on we will set $c=d=1$, as is common in the 
literature \citep{Prigogine71}. In the deterministic case, the linear stability 
analysis about this fixed point is a special case of that carried out above 
in the stochastic picture, and corresponds to ignoring the noise term in 
Eq.~\eqref{Fourier_linearised}. Therefore the small, deterministic, spatially 
inhomogeneous, perturbations $\tilde{\xi}_{\gamma}(k, \tau)$ satisfy the
equation
\begin{equation}
\frac{\partial \tilde{\xi}_{\gamma}(k, \tau)}{\partial \tau} = 
\sum_{\delta = 1}^{2}{\mathcal J}^{*}_{\gamma \delta}(k)\,\tilde{\xi}_{\gamma}
(k, \tau),
\label{LSA}
\end{equation}
where the Jacobian is found to be 
\begin{equation}
{\mathcal J}^{*}(k) = \left( 
\begin{array}{cc}
b - 1 + \alpha \tilde{\Delta}_{k} & a^2 \\
- b & - a^2 + \beta \tilde{\Delta}_{k}
\end{array} 
\right).
\label{Jacobian} 
\end{equation}

The eigenvalues of the Jacobian, $\lambda_{\gamma}(k)$ ($\gamma=1,2$), give 
information about the stability of the homogeneous state. In particular, the 
perturbations $\tilde{\xi}_{\gamma}(k, \tau)$ grow like linear combinations of 
$e^{\lambda_{\gamma}(k)\,t}$, therefore if $\mbox{Re}[\lambda_{\gamma}(k)]$ is 
positive for some $k$ and some $\gamma$, then the  perturbation will grow with
time and the homogeneous state will be unstable for this value of $k$. Turing's
insight was that the pattern eventually formed as a result of the perturbation
is characterised by this value of $k$. The overall scenario is complicated by 
the nature of the boundary conditions, the presence of other attractors and 
the effect of the non-linearities \citep{Cross09}, but in the following we 
shall ignore these, and consider only the simplest case in order to understand 
the main concepts.

In this most straightforward situation, the small perturbation which excites 
the unstable $k$-th Fourier mode will cause the concentrations $u$ and $v$ 
to develop a sinusoidal profile about their fixed point values characterised 
by the wave-number $k$. The pattern is steady or pulsating depending on whether 
or not the imaginary part $\mbox{Im}[\lambda_{\gamma}(k)]$ is zero. In both 
cases, the amplitude of sinusoidal profile increases exponentially with a 
time-scale $1/\mbox{Re}[\lambda_{\gamma}(k)]$, and so clearly the eigenvalue
with the largest real part will dominate. By moving away from the homogeneous 
state the linear approximation will eventually lose its validity and the 
effect of the non-linear terms will become relevant. If the system admits no 
solutions which diverge, the growth will be damped by the non-linearities to 
some non-zero value, which defines the final amplitude of the spatial pattern.

\begin{figure}[t]
\begin{center}
\includegraphics[scale=0.95]{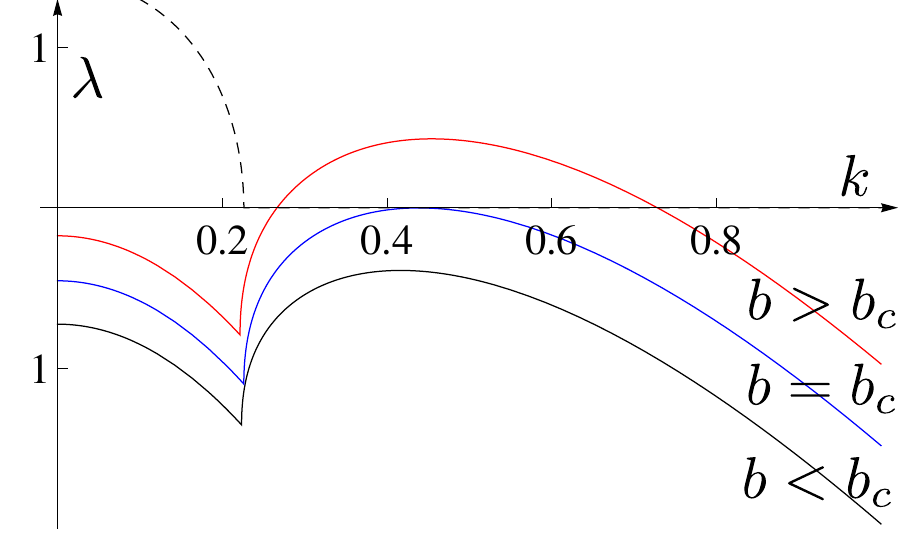} \\
\caption{Real part (solid lines) and imaginary part (dashed line) of the most
unstable eigenvalue of matrix $\mathcal J$. Parameter values are $a=1.5$,
$\alpha =2.8$ and $\beta = 22.4$, \citep{Cross09}. Solid lines correspond to
$b=1.8$ (black), $b\equiv b_{c}=2.34$ (blue) and $b=2.9$ (red). The imaginary
part is shown for $b=2.34$ only, although it looks qualitatively the same
for the range of $b$ values displayed.}
\label{fig:bruss-eigenv}
\end{center}
\end{figure}

Typically, the interesting case occurs when a control parameter triggers the 
pattern formation by making the real part of one of the eigenvalues positive. 
For the Brusselator this is illustrated in Fig.~\ref{fig:bruss-eigenv}, where 
the relevant eigenvalue of $\mathcal J$, (i.e.~the one which becomes positive) 
is shown for different values of the parameter $b$. Here $b$ is the control 
parameter with the other free parameter $a$ fixed and equal to $1.5$. 
For $b<b_{c}\approx 2.34$, the real part of both eigenvalues is negative and 
thus the homogeneous state is stable. This corresponds to the situation where 
there are no patterns. The critical value of $b$, $b_{c}$, occurs when the real
part of one of the eigenvalues is tangent to the $k$-axis. For values of $b$ 
larger than $b_{c}$, a window of unstable modes sets in, delimited by the 
intersections of $\mbox{Re}[\lambda]$ with the $k$-axis. Each mode contributes 
to the pattern, although the wavelength which maximises $\mbox{Re}[\lambda]$ 
is the one with bigger amplitude as it grows faster than the other modes.   


\begin{figure}[t]
\begin{center}
\includegraphics[scale=0.8]{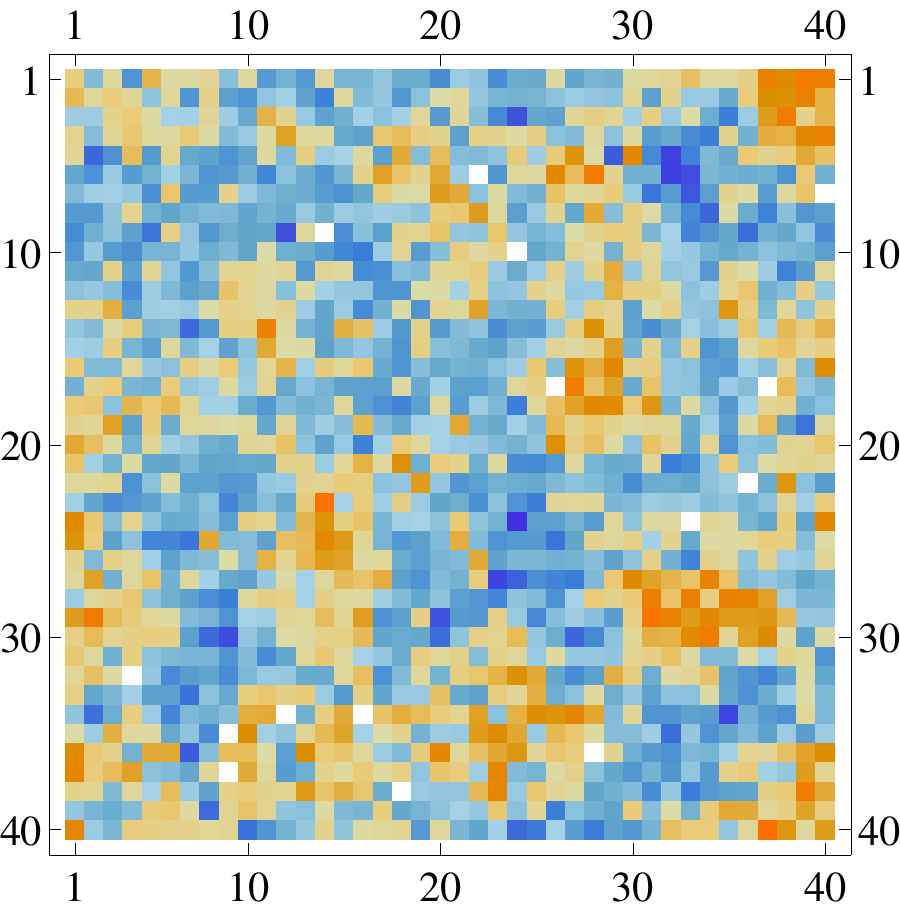} \\
\caption{(Colour online) Snapshot of two-dimensional stochastic Turing
patterns for species $X$. The system consists of $40\times40$ domains
with periodic boundary conditions. The parameters are the same as in
Fig.~\ref{fig:bruss-eigenv}, except that $b=2$ and $V=500$. Simulations
started close to $(u^*, v^*)$ and ran for $t/V = 15$. We have used
warm colours for values of $u>u^*$ and cold colours for $u<u^*$. White
pixels indicate the fixed point value, $u^*$. }
\label{fig:2d-patterns}
\end{center}
\end{figure}

As we have mentioned already no Turing patterns emerge from the deterministic 
equations if both eigenvalues of ${\mathcal J^{*}}$ have a negative real part 
for all $k$, since then the perturbations decay away. However, when the 
system is described as an IBM, intrinsic noise is present which acts as a 
continuous perturbation on the homogeneous state, exciting each $k$-mode. Given
that the homogeneous state is stable for all $k$, every excitation decays,
although each with a different time-scale given by 
$1/\mbox{Re}[\lambda_{\gamma}(k)]$. Thus, the closer 
$\mbox{Re}[\lambda_{\gamma}(k)]$ is to zero, the slower the relaxation of the 
corresponding $k$-mode. The situation is illustrated in 
Fig.~\ref{fig:bruss-eigenv}, where the value of $k$ at which this occurs for 
$b=1.8<b_c$ is seen to be the maximum of the curve of 
$\mbox{Re}[\lambda_{\gamma}(k)]$ versus $k$. The net effect is that only a 
window of modes around the maximum of $\mbox{Re}[\lambda_{\gamma}(k)]$ is 
visible in the dynamics, the others having died away. This patterns have been
called stochastic Turing patterns \citep{Biancalani10}. 
Figure \ref{fig:2d-patterns} shows the result of numerically simulating a 
two-dimensional Brusselator model using the Gillespie algorithm in a parameter 
regime where the homogeneous state is stable to perturbations for all $k$.


From the argument given above we would expect that in order to study stochastic 
Turing patterns it would be sufficient to linearise about the homogeneous 
state as before, but now to include noise. In addition, we would expect that
the detailed nature of the noise would be unimportant, only its strength. This
justifies the use of the LNA in the analysis of stochastic Turing patterns and
implies that the arguments which we have used can be made quantitative through
the use of Eq.~(\ref{Fourier_linearised}). To do this we first take the 
temporal Fourier transform of this equation to obtain 
\begin{equation}
\tilde{\xi}_{\gamma}(k, \omega) = 
\sum_{\delta,\sigma}\,\Phi^{-1}_{\gamma \delta}(k,\omega)\,g^{*}_{\delta
\sigma}(k)\,
\tilde{\eta}_{\sigma}(k,\omega),
\label{xi}
\end{equation}
where $\Phi_{\gamma \delta}(k,\omega) = -i\omega\delta_{\gamma \delta} -
{\mathcal J}^{*}_{\gamma \delta}$. From Eq.~(\ref{xi}) we can find an expression
for the 
power spectrum of the fluctuations which is the quantity we use to analyse 
the patterns:
\begin{equation} \label{PS}
P_\gamma (k, \omega ) = \langle |\tilde{\xi}_\gamma(k, \omega)|^2 \rangle 
= \sum_{\delta,\sigma}\,\Phi^{-1}_{\gamma \delta}(k,\omega)\,
\tilde{B}^{*}_{\delta \sigma}(k)\,(\Phi^{\dag})^{-1}_{\sigma \gamma}(k,\omega),
\end{equation}
where $\tilde{B}^{*}(k)$ is obtained by Fourier transforming in space the 
matrix $B(\boldsymbol y)$ given by Eq.~\eqref{B_ggT}, evaluated at the fixed 
point. The details of this calculation can be found in \cite{Lugo08}; here we
simply state the final formulae --- which holds for any two-species system 
which has one spatial dimension:
\begin{eqnarray} \label{B_ft}
\tilde{B}^*_{11}(k)= B^*_{11} - 2 u^* \alpha \tilde{\Delta}_{k}, \nonumber \\
\tilde{B}^*_{12}(k) = B^*_{12}, \quad \tilde{B}^*_{12}(k) = B^*_{12}, \nonumber
\\
\tilde{B}^*_{22}(k)= B^*_{22} - 2 v^* \beta \tilde{\Delta}_{ k}.
\end{eqnarray}
Here, the matrix $B^*$ indicates the correlation matrix of Eq.~\eqref{B_ggT} 
calculated at the fixed point for the corresponding non-spatial system. For 
instance, in the case of the Brusselator this is obtained by considering only 
the reactions \eqref{Bruss_reactions_loc} without those of 
Eqs.~\eqref{Bruss_reactions_mig}, which yields: $B^*_{11} = 2 a (1 + b)$,
$B^*_{12} = B^*_{21} = - 2 a b$ and $B^*_{22} = 2 a b$.


\begin{figure}[t]
\begin{center}
\includegraphics[scale=0.6]{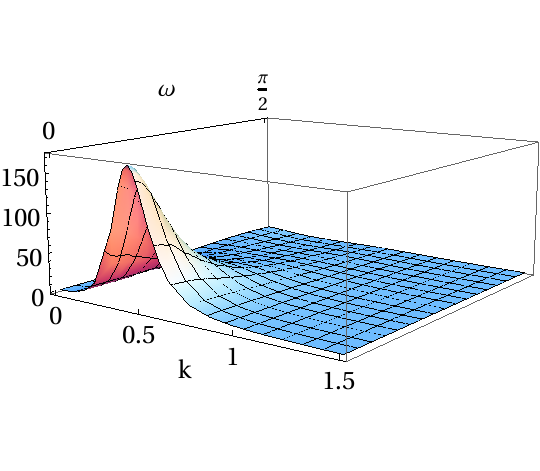}
\caption{(Colour online) Analytical power spectrum of species $X$, obtained 
with the same parameter values as in Fig.~\ref{fig:2d-patterns}, and from the
analytical expression \eqref{PS}.} 
\label{fig:anal-ps}
\end{center}
\end{figure}



The expression \eqref{PS} for the power is plotted in Fig.~\ref{fig:anal-ps}. 
We have also measured the numerical power spectrum via the Gillespie algorithm,
and found good agreement with the analytical expression, confirming that the 
dynamics is captured within the approximation scheme we have used. The 
power spectrum shows a peak at $k \neq 0$ and $\omega = 0$ which indicates the 
presence of stochastic Turing patterns of length scale characterised by $k$. 
As shown in previous studies \citep{Butler09, Biancalani10}, one can compute 
the region of parameters for which stochastic Turing patterns arise by looking 
at when $P_\gamma (k, 0)$ has a maximum for some non-zero $k$. It has been 
found that those regions are greatly enlarged, making the pattern formation a 
much more robust mechanism. Specifically, stochastic patterns may appear even 
for equal diffusivities, a condition for which deterministic patterns cannot 
occur \citep{Butler09,Biancalani10,Butler11}.

\begin{figure}[t]
\begin{center}
\includegraphics[scale=1.1]{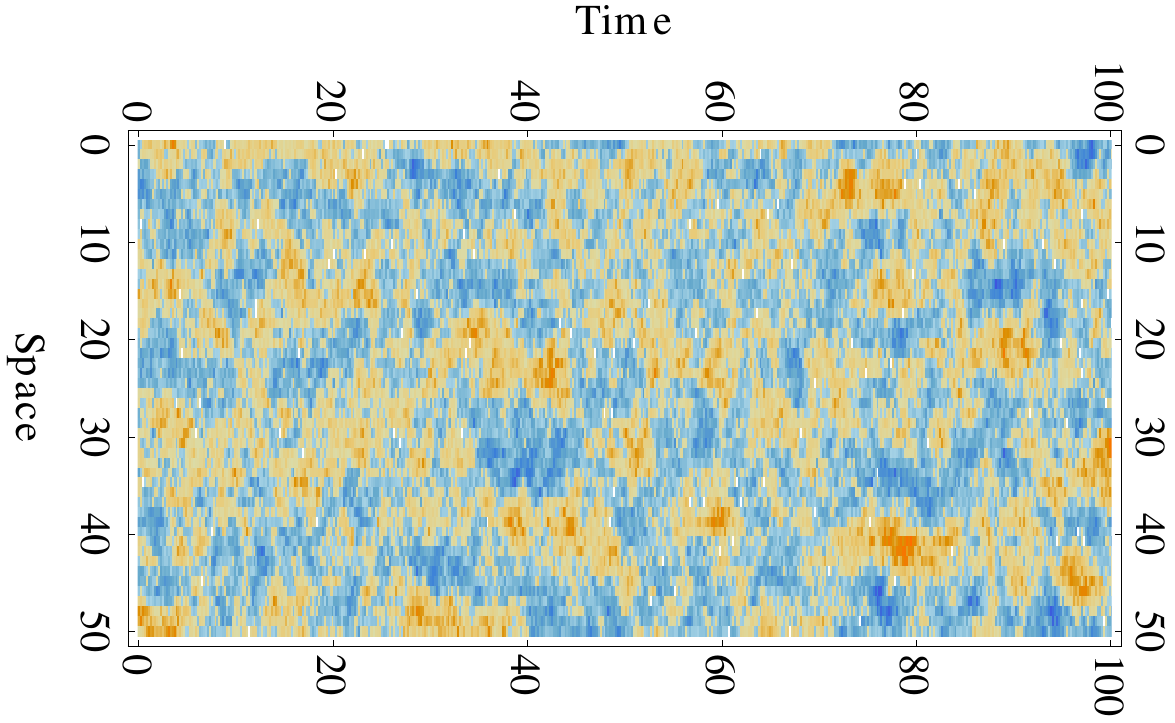} \\
\caption{(Colour online) Dynamics of a one-dimensional system of $50$ domains 
run for $2*10^3$ time ($\tau = t/V$) units. Parameter values are the same 
as in  Fig.~\ref{fig:2d-patterns}.}
\label{fig:1d-pattern}
\end{center}
\end{figure}

Notice that unlike their deterministic analogue, stochastic patterns are not 
steady but they continuously decay whilst they are re-created by the effect of 
the noise \citep{Scott11,Biancalani11}. In Fig.~\ref{fig:1d-pattern} this 
effect is shown by means of the dynamics of a one-dimensional Brusselator. The 
noisy nature of patterns makes them hard to detect by the naked eye, and the 
emergence of a length scale only becomes clear by means of a Fourier analysis.


The theory we have presented here is rather general and also applies to 
other types of pattern instabilities. For instance, if the power spectrum 
showed a peak at $k \neq 0$ and $\omega \neq 0$ the overall pattern would 
consist of stochastic travelling waves \citep{Biancalani11}. Finally, it should 
be mentioned that the amplitude of stochastic patterns scales as $1/\sqrt{V}$ 
and it therefore vanishes in the limit $V\rightarrow \infty$, in which the 
deterministic picture is recovered. Stochastic patterns arise because of the 
noise related to the discreteness of the populations and they are therefore 
less relevant for populations in which the number of individuals is 
macroscopic. 

\section{Spontaneous Emergence of Cell Polarity}     
\label{cell_polarity}
Many important biological functions require cells to break their rotational
symmetry and form a distinguished `nose' and `tail'. The emergence of this
symmetry-breaking is known as polarisation, and the precise mechanisms
responsible are not yet fully understood. A few years ago \cite{Altschuler08} 
proposed a very simple model of polarisation in which signalling molecules 
self-recruit to the cell membrane, before diffusing away. They showed through 
simulations that, depending on the choice of model parameters, the membrane 
molecules may spontaneously  aggregate. 

Further investigations have followed several different lines, including more
detailed simulations \citep{Petzold} and mathematically rigorous studies
\citep{Gupta12}. In this section, we will show how the mesoscopic framework
developed in previous sections may be used to analytically describe the
spontaneous emergence of cell polarity in this model. This effect is stronger
than those described by the LNA, and it will require a different theoretical
approach.

Before beginning the analysis, it is worth noticing that in this model the total
number of molecules does not change; there is thus no need to distinguish
between this and the volume, so we write $N=V$. Moreover, the variables $\ell$
(giving the number of molecules in the cytoplasmic pool) and $m_i$ (the number
of molecules in membrane domain $i$) are related by: $\ell+\sum_im_i=V\,.$

As usual, we first explore the behaviour of the macroscopic equations.
Substituting the transition rates (\ref{polar_reactions_loc}) and
(\ref{polar_reactions_mig}) into equations (\ref{macroscopic_eqn}) and
(\ref{f_defn}), we arrive at 
\begin{eqnarray}
\frac{du}{d\tau}&=&(\koff-u\,\kfb)\sum_iv_i-u\,\kon\nonumber\\
\frac{dv_i}{d\tau}&=&u\,\kon +(u\,\kfb-\koff)\,v_i+\alpha\Delta v_i\nonumber\,.
\end{eqnarray}
Conservation of the total number of molecules implies that $u+\sum_jv_j=1$, so
we may eliminate $u$ from the above system to obtain 
\begin{equation}
\frac{dv_i}{d\tau}=\Big(1-\sum_jv_j\Big)\big(\kon+v_i\,\kfb\big)
-v_i\koff+\alpha\Delta v_i\,.
\label{polar_macro}
\end{equation}
In the appropriate continuum limit, this equation agrees with that of 
\cite{Altschuler08}. 

As with the Brusselator, there is a homogeneous fixed point. Putting $v_i\equiv
v$ and setting $dv_i/d\tau=0$, we find the quadratic equation
\begin{equation}
(1-\Omega\,v)(\kon+v\,\kfb)-v\,\koff=0\,.
\end{equation}
For simplicity, we consider the case in which $\kon\approx0$ (that is, almost
all membrane molecules exist as result of the feedback mechanism). In this
limit, the homogeneous fixed point is given by $v=v^{*}/\Omega$, where $v^*$ is
the mean fraction of molecules on the membrane:
\begin{equation}
v^{*}=\begin{cases}\displaystyle\left(1-\frac{\koff}{\kfb}\right)\quad&\text{if}
\quad\kfb>\koff\\0&\text{otherwise.}\end{cases}
\end{equation}
To gain further insight, we pass to Fourier space, where Eq.~(\ref{polar_macro})
with $\kon=0$ becomes
\begin{equation}
\frac{d\tilde{v}_k}{dt}=\tilde{v}_k\Big[\kfb(1- l^{-1}\tilde{v}_0)
-\koff+\alpha\big(\cos( l k)-1\big)\Big]\,.
\label{polar_macro_f}
\end{equation}
The Jacobian for this system is diagonal, and it is straightforward to read off
the eigenvalues at the fixed point $\tilde{v}_k=\delta_{w,0}\, l\,v^*$ as 
$$\lambda_k=\begin{cases}\koff-\kfb\Big.\,,\quad&\text{if}\quad
k=0\\\Big.\alpha\big(\cos( l k)-1\big)&\text{if}\quad k\neq 0\,.\end{cases}$$
We can conclude from this analysis that provided $\kfb>\koff$ the homogeneous
fixed point is non-zero and stable. This is a puzzle: the homogeneous state
corresponds to the signalling molecules being spread uniformly around the
membrane, if this state is stable, then how can polarisation occur?

We postulate that the answer lies in the following observation. Notice that if
the diffusion coefficient $\alpha$ is small then the modes with wave number
$k\neq0$ are only marginally stable; their associated eigenvalues are close to
zero. In this regime, a small random perturbation (resulting from intrinsic
noise, for example) may be enough to push the system very far from its
equilibrium state. Moreover, the stochastic dynamics in this regime cannot be
understood within the framework of the LNA, for the simple reason that when the
system has been pushed far from its steady state, linearisation around that
state is no longer representative of the true dynamics. To make analytical
progress, we will need to deal with the non-linearity of the model some other
way. 

We begin by writing down the mesoscopic equations. For our purposes, the
Fokker-Planck equation (\ref{FPE}) is the most useful, with $A$ and $B$ given by
\begin{eqnarray}
A_i(\bm{v})&=&\Big(1-\sum_jv_j\Big)\big(\kon+v_i\,\kfb\big)
-v_i\koff+\alpha\Delta v_i\,,\nonumber\\
B_{ij}(\bm{v})&=&\delta_{ij}\bigg[\Big(1-\sum_k v_k\Big)\big(\kon+v_i\,\kfb\big)
+v_i\koff\bigg]\,.\nonumber
\end{eqnarray}
Note that we have neglected terms of order $\alpha/V$ from the noise, as we are
interested in behaviour when $\alpha$ is small and $V$ large, so $\alpha/V$ is
negligible. As with the macroscopic equations, this system is easier to analyse
in Fourier space. We introduce the distribution $P(\bm{\tilde{v}},\tau)$ of 
Fourier variables $\tilde{v}_k$, which satisfies the Fokker-Planck equation
\begin{eqnarray}
\frac{\partial P(\bm{\tilde{v}},\tau)}{\partial \tau} &=& 
- \sum_{k} \frac{\partial }{\partial \tilde{v}_k} 
\left[ \tilde{A}_{k}(\bm{\tilde{v}}) P(\bm{\tilde{v}},\tau) \right]  \\
&+& \frac{1}{2V} \sum_{k,k'}
\frac{\partial}{\partial\tilde{v}_k}\frac{\partial}{\partial
\overline{\tilde{v}_{k'}}} 
\left[ \tilde{B}_{k,k'}(\bm{\tilde{v}}) P(\bm{\tilde{v}},\tau)
\right]\,,\nonumber 
\end{eqnarray}
where
\begin{eqnarray}
\tilde{A}_{k}(\bm{\tilde{v}})&=&\tilde{v}_k\,\Big[\kfb(1-
l^{-1}\tilde{v}_0)-\koff+\alpha\big(\cos( l k)-1\big)\Big]\nonumber\\
\tilde{B}_{k,k'}(\bm{\tilde{v}})&=&\tilde{v}_{k+k'}\, l\,\Big[\kfb(1-
l^{-1}\tilde{v}_0)+\koff\Big]\,.
\label{AB_fourier}
\end{eqnarray}
Note that the Fourier modes may take complex values, and we use
$\partial/\partial \overline{\tilde{v}_{k'}}$ to denote differentiating with
respect to the complex conjugate. For later convenience, we assume $\Omega$ is
odd and number the modes by $k\in\{-(\Omega-1)/2,\ldots, (\Omega-1)/2\}$, so
that $\tilde{v}_{-k}=\overline{\tilde{v}_k}$.

Our remaining analysis is informed by two observations. First, we note that the
non-linearity in equation (\ref{AB_fourier}) arises only from terms involving
$\tilde{v}_0$. Second, in the interesting regime $\kfb-\koff\gg\alpha$, we have
that the eigenvalues of the macroscopic system satisfy $\lambda_0\ll\lambda_k<0$
and thus $\tilde{v}_0$ is (comparatively) very stable near $v^*$. This implies a
separation of time-scales in the problem: we expect $\tilde{v}_0$ to relax very
quickly to a value near its equilibrium, whilst the other modes fluctuate
stochastically on much slower time-scales. Combining these facts suggests the
following strategy: we restrict our attention to only those trajectories in
which $\tilde{v}_0$ is held constant at $l v^*$. 

Conditioning on the value of $\tilde{v}_0$ alters the structure of the noise
correlation matrix for the other modes; details of the general formulation are
given in Appendix B of \cite{Rogers12a}. The result is a Fokker-Planck equation
for the distribution $P^{(c)}(\bm{\tilde{v}},\tau)$ of the remaining Fourier
modes $\tilde{v}_k$ with $k\neq0$, conditioned on $\tilde{v}_0$ taking the value
$ lv^*$:
\begin{eqnarray}
\frac{\partial P^{(c)}(\bm{\tilde{v}},\tau)}{\partial \tau} &=& 
- \sum_{ki\neq0} \frac{\partial }{\partial \tilde{v}_i} 
\left[ \tilde{A}^{(c)}_{k}(\bm{\tilde{v}}) P(\bm{\tilde{v}},\tau) \right]  \\
&+& \frac{1}{2V} \sum_{k,k'\neq0}
\frac{\partial}{\partial\tilde{v}_k}\frac{\partial}{\partial
\overline{\tilde{v}_{k'}}} 
\left[ \tilde{B}^{(c)}_{k,k'}(\bm{\tilde{v}}) P(\bm{\tilde{v}},\tau)
\right]\,,\nonumber
\label{FPEc}
\end{eqnarray}
where
\begin{eqnarray}
\tilde{A}^{(c)}_{k}(\bm{\tilde{v}})&=&\alpha\,\tilde{v}_k\,\big(\cos( l
k)-1\big)\bigg.\nonumber\\
\tilde{B}^{(c)}_{k,k'}(\bm{\tilde{v}})&=&2\koff\bigg[
l\,\tilde{v}_{k+k'}-\frac{\tilde{v}_{k}\tilde{v}_{k'}}{v^*}\bigg]\,.
\end{eqnarray}
Multiplying (\ref{FPEc}) by $\tilde{v}_k$ and integrating over all
$\bm{\tilde{v}}$ we obtain a differential equation for the mode averages:
\begin{equation}
\frac{d}{d\tau}\langle\tilde{v}_k\rangle=\alpha\langle\tilde{v}
_k\rangle\big(\cos( l k)-1\big)\,.
\end{equation}
Thus, for all $\alpha>0$ we have
\begin{equation}
\langle\tilde{v}_k\rangle\to\delta_{k,0}\, lv^*\,.
\end{equation}
For the second-order moments the behaviour is not so trivial. Multiplying
(\ref{FPEc}) this time by $\tilde{v}_k\tilde{v}_{k'}$ and integrating yields
\begin{equation}
\frac{d}{d\tau}\langle\tilde{v}_k\tilde{v}_{k'}\rangle=\frac{l\koff}{V}
\langle\tilde{v}_{k+k'}\rangle
+\langle\tilde{v}_k\tilde{v}_{k'}\rangle\Big[\alpha\big(\cos( l k)+\cos( l
k')-2\big)-\frac{1}{V}\frac{\koff}{v^*}\Big]\,.
\end{equation}
The equilibrium values are thus $\langle\tilde{v}_k\tilde{v}_{k'}\rangle\to0$ if
$k+k'\neq0\mod\Omega$, and 
\begin{equation}
\langle|\tilde{v}_k|^2\rangle\to\frac{(lv^*)^2}{1+\alpha V
(1-\cos(lk))v^*/\koff}\,.
\label{fourier_magnitude}
\end{equation}
This is our main result, although further analysis is required to interpret the
implications for the behaviour of the model. 

\begin{figure}[t]
\begin{center}
\includegraphics[width=0.5\textwidth, trim=20 280 100 280]{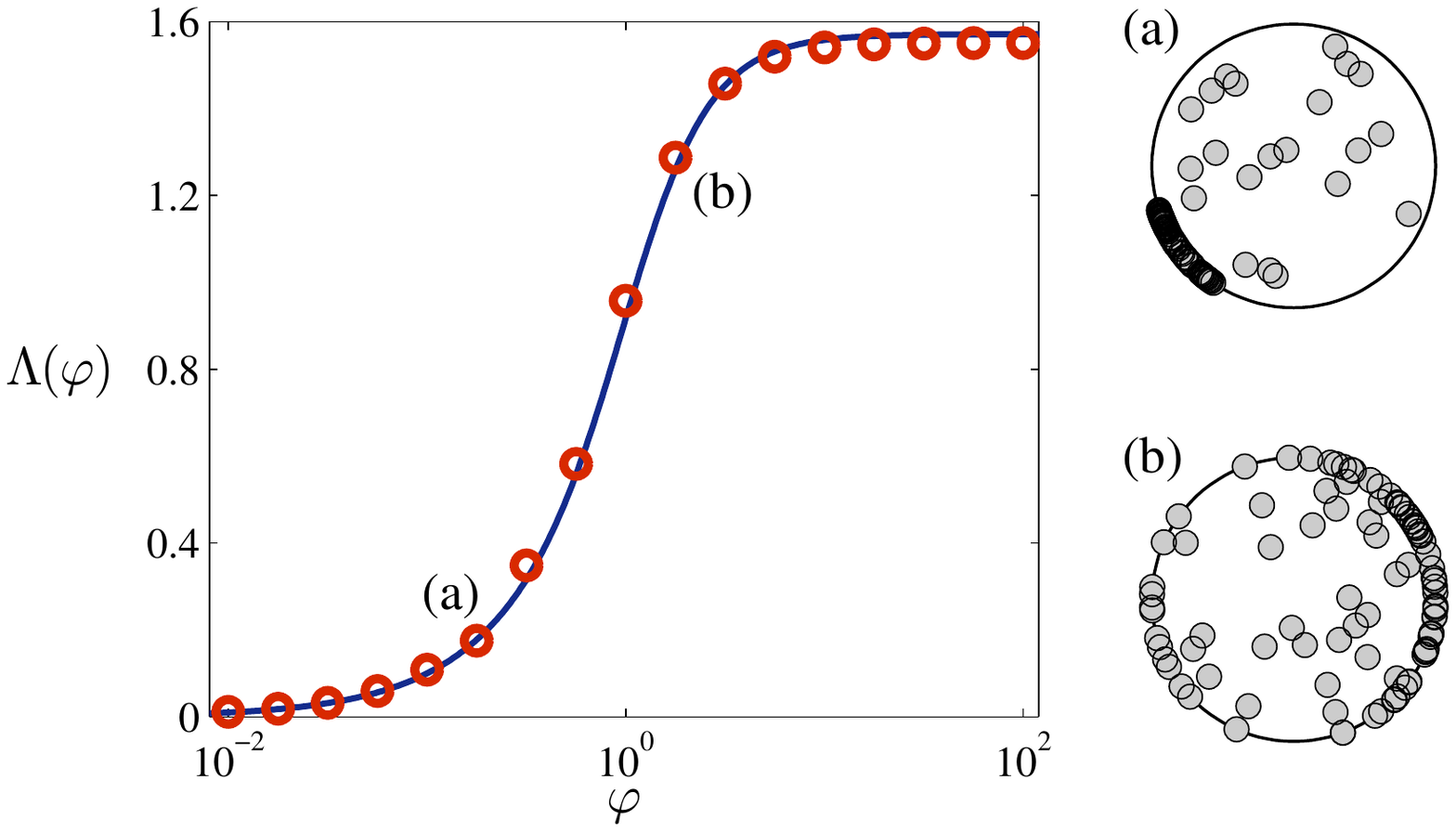}
\caption{(Colour online) Mean angular distance between membrane molecules, 
as $\varphi$ is varied over several orders of magnitude. Red circles are the 
results of simulations, the blue line shows the result from Eq.~(\ref{Lambda}).
On the right are two snapshots from simulations, with $\varphi$ corresponding 
to the points (a) and (b) in the main figure. Clearly (a) is polarised and 
(b) is not, as predicted by the theory.}
\label{fig:lambda}
\end{center}
\end{figure}

In \cite{Altschuler08} it was observed that simulations of a continuum version
of this model exhibited the curious phenomenon of the membrane molecules
grouping together, despite the macroscopic equations suggesting they should be
spread uniformly around the membrane. We introduce a summary statistic to
measure this effect. Suppose the membrane molecules are distributed according to
an angular density field $v(x)$, and let $\Lambda$ denote the mean angular
separation of two molecules:
\begin{equation}
\Lambda=\left(\frac{1}{v^*}\right)^2\int_{-\pi}^\pi\int_{-\pi}^\pi
d(x,y)\big\langle v(x)v(y)\big\rangle\,dx\,dy\,,
\end{equation}
where 
\begin{equation}
d(x,y)=\begin{cases}|x-y|\quad&\text{if}\quad
|x-y|<\pi\\2\pi-|x-y|&\text{otherwise.} \end{cases}
\end{equation}
To compute $\Lambda$ from our result (\ref{fourier_magnitude}) we pass to the
continuum limit, taking $\Omega\to\infty$ and using the modes $\tilde{v}_k$ as
the coefficients of the Fourier series of the membrane angular density field
$v(x)$. Taking an angular prescription for the membrane so that $l=2\pi/\Omega$,
we renormalise by a factor of $1/2\pi l$ and reverse the Fourier transform to
obtain 
\begin{equation}
v(x)=\lim_{\Omega\to\infty}\frac{1}{2\pi l}\sum_{k=1}^\Omega
e^{ikx}\tilde{v}_k\,,\quad\text{for }x\in[-\pi,\pi)\,.
\label{vx}
\end{equation}
The calculation proceeds thus:
\begin{eqnarray}
\Lambda&=&2\pi\left(\frac{1}{v^*}\right)^2 \int_{-\pi}^\pi|x|\big\langle
v(x)v(0)\big\rangle\,dx\nonumber\\
&=&\frac{1}{2\pi}\left(\frac{1}{lv^*}\right)^2\sum_{k,k'}
\langle\tilde{v}_k\tilde{v}_{k'} \rangle
\int_{-\pi}^\pi|x|e^{ikx}\,dx\nonumber\\
&=&\frac{\pi}{2}+\sum_{k\neq0} \frac{1}{1+\varphi^2
k^2}\int_{-\pi}^\pi|x|e^{ikx}\,dx\,,
\end{eqnarray}
where $\varphi=(\alpha V l^2  v^* /\koff)^{1/2}$, which we assume to have a
finite value in the limit $l\to0$. The first equality above comes from the
rotational invariance of the model meaning that we may fix $y=0$, after which we
employ Eq.~(\ref{vx}) and Eq.~(\ref{fourier_magnitude}) in turn. Now, for
$k\neq0$ 
\begin{equation*}
\int_{-\pi}^\pi|x|e^{ikx}\,dx=\begin{cases}-4k^{-2}\quad&\text{for odd $k$}\\0
&\text{for even $k$.}\end{cases}
\end{equation*}
Also, the following infinite series \citep{Bromwich1926} will be useful: 
\begin{equation*}
\sum_{k\,\text{odd}}\frac{1}{x^2+k^2}=\frac{\pi }{2x}\tanh\left(\frac{\pi
x}{2}\right)\,.
\end{equation*}
Altogether, we have
\begin{eqnarray}
\Lambda(\varphi)&=&\frac{\pi}{2} - \frac{2}{\pi}\sum_{k\,\text{odd}}
\left(\frac{1}{1+\varphi^2 k^2}\right)\left(\frac{1}{k^2}\right)\nonumber\\
&=&\frac{\pi}{2} -
\frac{2}{\pi}\left(\,\sum_{k\,\text{odd}}\frac{1}{k^2}-\sum_{k\,\text{odd}}\frac
{1}{\varphi^{-2}+k^2}\right)\nonumber\\&=&\Bigg.\varphi\tanh\left(\frac{\pi}{
2\varphi}\right)\,.
\label{Lambda}
\end{eqnarray}
The limits are $\Lambda(\infty)=\pi/2$, which corresponds to molecules spread
uniformly around the membrane, and $\Lambda(0)=0$, which corresponds to complete
localisation. In figure~\ref{fig:lambda} we compare this theoretical prediction
with the result of simulations for $\varphi$ varying over several orders of
magnitude. For small values of $\varphi$ the membrane molecules cluster into a
tight group, meaning that the cell has become polarised. As $\varphi$ is
increased (caused by an over-abundance of signalling molecules in relation to
their rate of diffusion around the membrane) this effect is weakened, and the
cell loses polarity. 


Finally, we discuss the physical meaning of the parameter $\varphi$. The average
number of molecules on the membrane at equilibrium is $V v^*$, the typical
lifetime of a membrane molecule is $1/\koff$, and the continuum diffusion
coefficient is $\alpha l^2$. The quantity $\varphi$ can thus be interpreted as
the mean total distance travelled by all the membrane molecules during their
lifetime: if this quantity is small, they must remain localised. 

\section{Discussion and Conclusion}
\label{conclude}
The main aim of this paper has been to show that the analysis of stochastic 
models in biology need not only be numerical: a range of analytical techniques
are available, just as for deterministic models. In fact the treatment of 
stochastic models may be simpler, since in many cases the noise can be 
considered to be a linear perturbation (the LNA) to the deterministic form of 
the dynamical equations. Linear equations such as these are easier to solve, 
especially when the fluctuations are around a stationary state.

The deterministic, or macroscopic, description of the system is valid when the 
individual nature of the constituents is unimportant, for example when they 
are so numerous as to effectively be infinite in number. The ensemble average
of the stochastic variables will also obey the same deterministic equation. 
The general form of this equation is Eq.~(\ref{macroscopic_eqn}), that is,
$\dot{y}_{I} = A_I(\boldsymbol{y})$, where the dot denotes a time derivative 
and the index $I$ includes both a position label and another identifying the 
constituent type. The function $A_I$ can be calculated from the elementary 
reactions of the process using Eq.~(\ref{A_I}). The mesoscopic, or stochastic,
description of the system which uses the same variables as the macroscopic 
description, is in principle no different. The general form of this equation
is (\ref{mesoscopic_eqn}), that is,
$\dot{y}_{I} = A_I(\boldsymbol{y}) +
V^{-1/2}\sum_{J}g_{IJ}(\boldsymbol{y})\eta_J$, where $\eta_J$ is a Gaussian
white noise with zero mean and unit strength. 
The only additional function which appears over and above that in the 
deterministic equation is $g_{IJ}$, which can, like $A_I$, be calculated from 
the elementary reactions of the process using Eqs.~(\ref{B_IJ}) and 
(\ref{B_ggT}). Although  Eq.~(\ref{mesoscopic_eqn}) is a straightforward 
generalisation of Eq.~(\ref{macroscopic_eqn}), it is much less well-known.

There are several reasons for the perceived difficulty of using  
Eq.~(\ref{mesoscopic_eqn}), probably the most important being the unfamiliarity
of many biologists with the general theory of stochastic processes. We have 
tried to show in this paper that the stochastic theory which is required need 
not be more complicated than that of dynamical systems theory, which is 
applicable to equations such as (\ref{macroscopic_eqn}). This is especially 
true if the LNA is a valid approximation for the particular system under study.
If the multiplicative nature of the noise cannot be neglected, as in section 
\ref{cell_polarity}, then care is required because of the singular nature of
white noise. However, even in this case, a systematic theory has been developed
that  may be applicable in situations in which there is a separation of 
time-scales \citep{Rogers12b,Biancalani12,Rogers12a}.

We applied Eq.~(\ref{mesoscopic_eqn}) to two sets of processes, one for which
the LNA was applicable and one for which it was not. The former situation was
discussed in section~\ref{bruss} where we revisited the problem of the 
emergence of spatial structures for systems of populations, in the paradigmatic
example of the Brusselator model. Intrinsic fluctuations, which are intuitively
thought of as a disturbing source, appear instead to be critical for the 
emergence of spatial order. More specifically, we showed how Turing patterns 
can arise for parameter values for which the macroscopic equations predict 
relaxation to a homogeneous state. We called these patterns `stochastic 
patterns', as they are generated by the continuous action of noise present in 
the system. However, it can be argued that the amplitude of stochastic patterns
might be so small that they can hardly be observed in a real population, given 
that the amplitude of the noise is small as well. Whilst this might be true 
for some systems, a recent study \citep{Ridolfi11} has suggested that the
response to a small perturbation in a pattern-forming system can be 
unexpectedly large, if the system displays a sufficient degree of 
`non-normality'. The connection between non-normality and stochastic patterns 
is so far largely unexplored, and constitutes a possible further investigation 
in this line of research.

In section~\ref{cell_polarity} we discussed an example of a stochastic 
phenomenon which goes beyond what can be understood within the LNA. The 
stochastic patterns appearing in the Brusselator are noise-driven perturbations
around the homogeneous state, having characteristic magnitude $1/\sqrt{V}$ (and
thus disappearing in the limit $V\to\infty$). By contrast, the spontaneous 
emergence of cell polarity in the model of Altschuler {\it et al} requires
the noise to have a more complex structure, which can lead the system to a 
state very far removed from the homogeneous fixed point of the deterministic 
equations. To characterise this process, it was necessary to study the full 
effect of the non-linear terms in the mesoscopic equations. To achieve this, 
we exploited the natural separation of time-scales occurring between the 
dynamics of the zeroth Fourier mode (which relaxes quickly to its equilibrium 
value) and the remaining degrees of freedom. This is a non-standard technique, 
however, it can be made relatively systematic and general, as will be outlined 
in a forthcoming paper. The LNA has played an important role in boosting the
recognition of the importance of stochastic effects  in the literature; we 
hope that methods employing the separation of time-scales may provide the 
next theoretical advance. 

\vspace{1cm}

{\bf Acknowledgements}. \ This work was supported in part under EPSRC Grant No. 
EP/H02171X/1 (A.J.M and T.R). T.B. also wishes to thank the EPSRC for partial 
support.


\end{document}